\documentclass[twocolumn,showpacs]{revtex4-1}

\usepackage{amsmath,bbm}
\usepackage{graphics}
\usepackage{epsfig}
\usepackage{epstopdf}
\usepackage{color}

\begin{document}

\title{Generation and characterization of discrete spatial entanglement\\in multimode nonlinear waveguides}

\author{Micha\l{} Jachura{$^{1}$}}
\email{michal.jachura@fuw.edu.pl}
\author{Micha\l{ } Karpi\'{n}ski{$^{1}$}}
\author{Konrad Banaszek{$^{1,2}$}}
\author{Divya Bharadwaj{$^{3}$}}
\author{Jasleen Lugani{$^{3}$}}
\author{K. Thyagarajan{$^{3}$}}

\affiliation{{$^{1}$}Faculty of Physics, University of Warsaw, Pasteura 5, 02-093 Warszawa, Poland}
\affiliation{{$^{2}$}Centre of New Technologies, Banacha 2c, 02-097 Warszawa, Poland}
\affiliation{{$^{3}$}Department of Physics, IIT Delhi, New Delhi 110016, India}

\begin{abstract}
We analyze theoretically spontaneous parametric down-conversion in a multimode nonlinear waveguide as a source of entangled pairs of spatial qubits,  realized as superpositions of a photon in two orthogonal transverse modes of the waveguide. It is shown that by exploiting intermodal dispersion, down-conversion into the relevant pairs of spatial modes can be selected by spectral filtering, which also provides means to fine-tune the properties of the generated entangled state. We also discuss an inverting interferometer detecting the spatial parity of the input beam as a versatile tool to characterize properties of the generated state. A single-photon Wigner function obtained by a scan of the displaced parity can be used to identify the basis modes of spatial qubit, whereas correlations between displaced parity measurements on two photons can directly verify quantum entanglement through a violation of Bell's inequalities.
\end{abstract}

\maketitle

\section{Introduction}

Optical systems find numerous applications in current efforts to implement emerging quantum enhanced technologies \cite{OBrien2009}. This prominence is owed in large measure to the availability of multiple degrees of freedom that permit a broad range of well-controlled manipulations on the generated quantum states of light \cite{Barreiro2005,Fickler2012,Brecht2015}. In particular, the spatial degree of freedom is currently studied in the context of quantum imaging and communication \cite{Dixon2012,Aspden2013,Vallone2014,Chrapkiewicz2015}. While the spatial characteristics of optical fields in free space and bulk media can be described by continuous variables, wave-guiding structures introduce a natural discrete set of transverse spatial modes. In classical optical communication they are becoming a valuable resource for data-multiplexing \cite{Bozinovic2013, Richardson2013} whereas in the field of quantum optics they can be naturally used to implement qubit- or qudit-based quantum protocols \cite{Voortesi2010,Leach2012}.

In this paper, we present a theoretical study of a multimode nonlinear waveguide as a medium to generate and charaterize spatial qubits. Such qubits are implemented as single photons prepared in superpositions of orthogonal transverse modes. In order to relate the studied scenario to existing manufacturing capabilities, we will consider a specific example of a multimode waveguide fabricated through an ion-exchange process in a periodically polled potassium titanyl phosphate (PPKTP) crystal \cite{Karlsson1997, Bierlein1987}. In recent years, large $\chi^{(2)}$ nonlinearities exhibited by such structures have been exploited to construct high-brightness down-conversion sources of photon pairs \cite{Zhong2009, Harder2016}. Although so far attention has been primarily focussed on generating photons in fundamental spatial modes \cite{Eckstein2011,Karpinski2012,Jachura2014}, it has been noted that by exploring higher-order modes one can access more complex forms of entanglement \cite{Saleh2009,Kang2014, Mosley2009, SPIE2012}. We will address here the feasibility of such an approach assuming realistic parameters of a PPKTP waveguide.

Specifically, we will analyze here generation of a maximally entangled state of two spatial qubits in orthogonal polarizations, each qubit spanned by a pair of fundamental and first-order transverse modes. We will show that such a state can be easily produced through a suitably arranged type-II down-conversion process in the waveguide subjected to spectral filtering. In addition, the spectral degree of freedom can be conveniently used to fine-tune the properties of the generated entangled state. We will also consider characterization of the produced two-photon state. We will show that the inverting interferometer \cite{Mukamel2003, Leary2009, Leach2004} is a very useful tool that enables identification of the transverse modes spanning the spatial qubits as well as verification of the generated entanglement. While qubit manipulations could be ultimately also integrated into waveguide structures, using e.g. electrooptic devices described in \cite{Bharadwaj2015}, the free-space interferometric approach allows one to test exclusively the generation stage. It can also provide complete information about the spatial characteristics of the produced state including the basis modes of the spatial qubit.

This paper is organized as follows. In Sec.~\ref{Sec:SpatialQubit} we describe the waveguide structure, discuss its transverse modes, and define spatial qubits. The down-conversion process in a multimode nonlinear waveguide is analyzed in detail in Sec.~\ref{WaveguideSPDC}. The wave function of photon pairs generated by selecting down-conversion into specific combinations of transverse modes is calculated in Sec.~\ref{Sec:TwoPhotonWaveFunction}. Sec.~\ref{Sec:Characterization} presents a method to identify basis modes of spatial qubits by measuring a single-photon Wigner function with the help of an inverting interferometer. The same measurement extended to photon pairs is applied in Sec.~\ref{Sec:Bell} to verify generated entanglement through the violation of Bell's inequalities. Finally, Sec.~\ref{Sec:Conclusions} concludes the paper.

\section{Waveguide modes and spatial qubits}
\label{Sec:SpatialQubit}
In our numerical simulations we assume realistic PPKTP waveguide dimensions of $w = 6~\mathrm{\mu m}$ width, $d = 10~\mathrm{\mu m}$ depth and $L = 1~\mathrm{mm}$ length. We use the crystallographic coordinate system with the waveguide propagation axis oriented along the $x$ direction and the non-linear medium occupying the area $z \le 0$ in the perpendicular plane. The waveguide is modelled with sharp boundaries in the $y$ direction and a diffusive refractive index profile along the $z$ direction given by:
\begin{equation}
n_{\kappa} = n_{\kappa,\mathrm{PPKTP}} + \Delta n_{\kappa} \mathrm{erfc}(-z/d), \quad \kappa=y,z,
\end{equation}
where $n_{\kappa,\mathrm{PPKTP}}$ is the refractive index of the substrate crystal for the light polarized along the $\kappa$-th axis and $\Delta n_{\kappa}$ denotes the maximal refractive index contrast of the waveguide. For the substrate refractive index we relied on Sellmeier equations parameters measured in \cite{Kato2002} whereas index contrasts $\Delta n_{y} = 0.021$ and $\Delta n_{z} = 0.025$  have been taken from waveguide manufacturer data (AdvR, USA). Waveguide modes were found numerically using the finite difference method described in \cite{Fallahkhair2008} assuming transverse electric (TE) propagation.

\begin{figure}
\centering
\includegraphics[width=7cm]{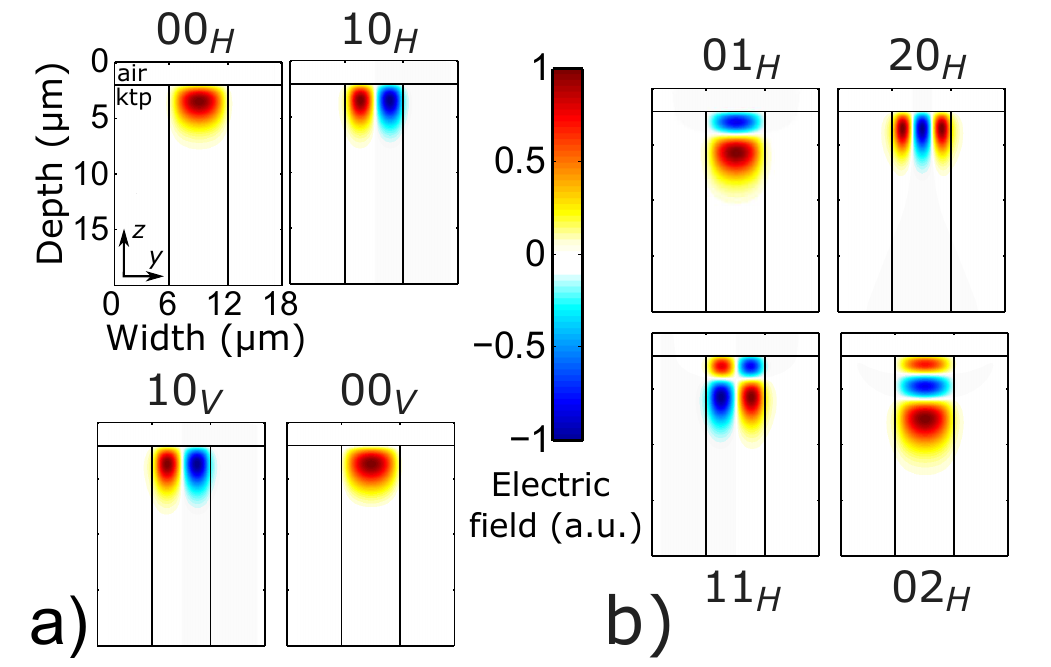}
\caption{(a) Waveguide modes involved in spatial entanglement generation. (b) Higher waveguide modes whose generation is suppressed by the means of intermodal-dispersion-based method.}
\label{Fig:WaveguideModes}
\end{figure}

Exemplary transverse modes at the 800~nm wavelength are depicted in Fig.~\ref{Fig:WaveguideModes} using respective electric field distributions $u^{ij}_{\mu}({\bf r})$ in the plane ${\bf r} = (y,z)$. The modes are labelled using two integers $ij$ specifying the number of nodes along the $y$ and $z$ direction respectively and a subscript $\mu=H, V$ denoting polarization. Altogether, the waveguide supports 12 modes for each of the two polarizations at the 770-830~nm wavelength range that will be used for the generation of photon pairs via the down-conversion process.
 
\begin{figure}[t]
\centering
\includegraphics[width=8cm]{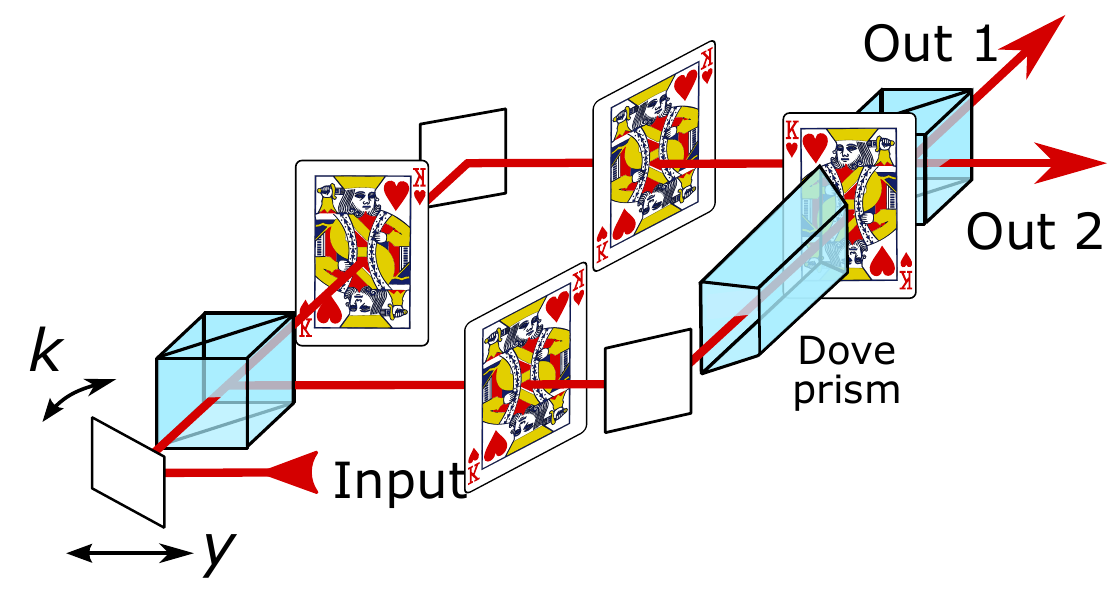}
\caption{Inverting Mach-Zehnder interferometer utilized for probing the parity of incoming spatial mode.}
\label{Fig:Interferometer}
\end{figure}

As the basis states for the spatial qubit, we will take a single photon prepared in transverse modes $00$ and $10$ shown in Fig.~\ref{Fig:WaveguideModes}(a). The waveguide can support two such qubits distinguished by photon polarization. The advantage of this specific choice is that the two modes used for the basis states have opposite parity with respect to the symmetry plane of the structure perpendicular to the waveguide facet. Consequently, they can be separated in a deterministic way using a Mach-Zehnder interferometer with an inverting Dove prism inserted in one of the arms, as shown in Fig.~\ref{Fig:Interferometer}. For the right choice of the relative phase between the interferometer arms, destructive interference occurs at a different output port of the interferometer depending on whether the input mode had an even or odd parity \cite{Leach2004,Leary2009,Mukamel2003}. Consequently, detecting the photon at the output ports of the interferometer implements a projective measurement in the spatial qubit basis. We will see in Sec.~\ref{Sec:Characterization} that by displacing and tilting laterally the input beam one can effectively determine the transverse modes that form the basis states of the spatial qubits.
Furthermore, sending the horizontally and vertically polarized photons to two such interferometers can be used to test the generated entanglement, as lateral displacement before the interferometer entrance provides a non-commuting measurement capable of violating Bell's inequalities \cite{Yarnall2007}. This idea will be discussed in Sec.~\ref{Sec:Bell}.

\section{Waveguide down-conversion}
\label{WaveguideSPDC}

\begin{figure}[b]
\centering
\includegraphics[width=8cm]{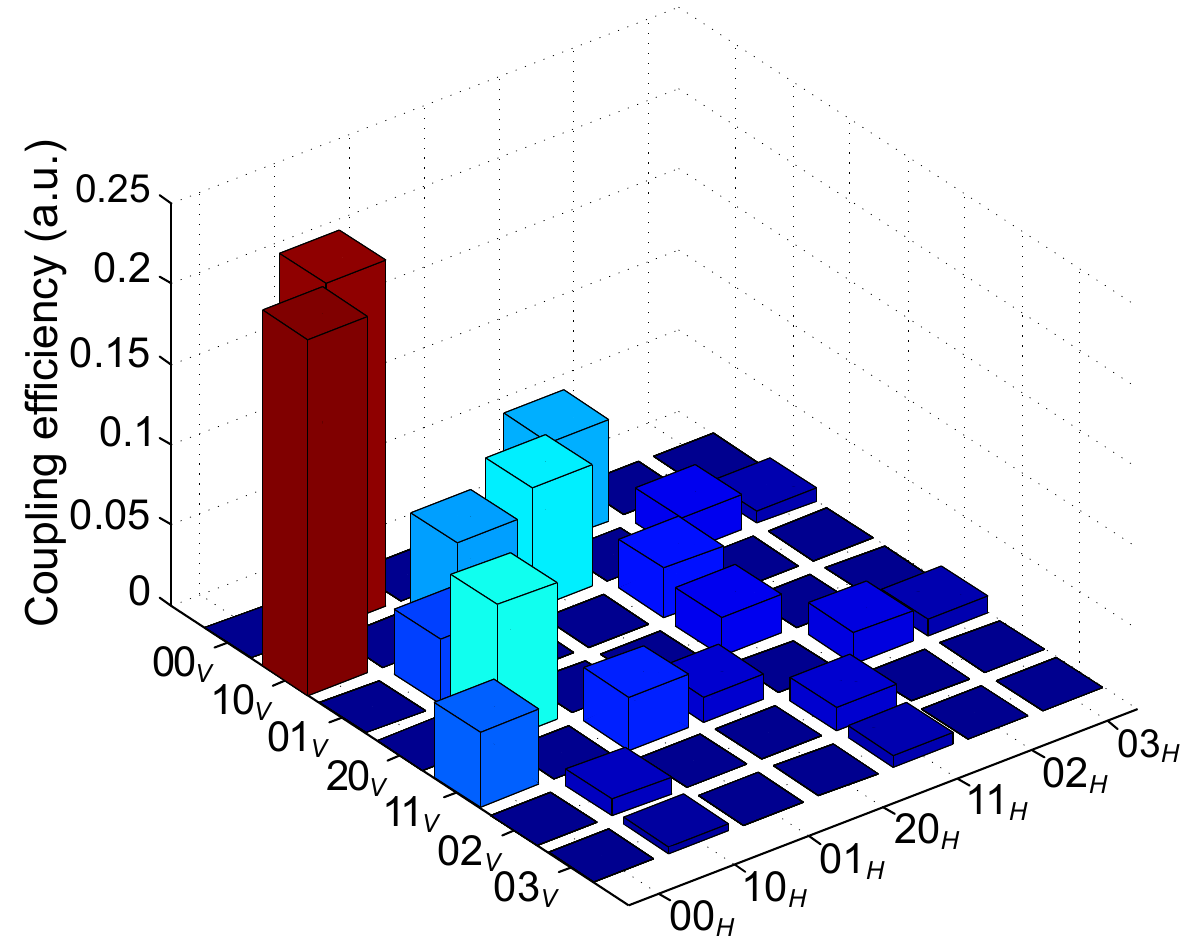}
\caption{Efficiencies of down-conversion processes for the pump coupled in $10_P$ mode $|\alpha_{10_{P} \rightarrow {ij}_{H} + {i'j'}_{V}}|^2$}
\label{Fig:CouplingEfficiencies10}

\end{figure}

Second-order $\chi^{(2)}$ nonlinearity of the PPKTP waveguide provides a possibility to generate entanglement between two spatial qubits described in Sec.~\ref{Sec:SpatialQubit}. This can be achieved by employing type-II spontaneous parametric down-conversion to produce a pair of photons distinguishable by their polarizations $H$ and $V$. In order to generate their entangled state, we will consider simultaneous realization of two down-conversion processes, producing pairs in modes $00_H 10_V$ and $10_H 00_V$. For experimental convenience, the pump $P$ should prepared in a single spatial mode. The efficiency of a specific down-conversion process $lm_{P} \rightarrow {ij}_{H}  {i'j'}_{V}$ depends on the spatial overlap of the three involved modes:
\begin{equation}
\alpha_{lm_{P} \rightarrow {ij}_{H}  {i'j'}_{V}} = \int d^2{\bf r} \, u_P^{lm} ({\bf r}) [u_H^{ij} ({\bf r}) u_V^{i'j'} ({\bf r})]^\ast.
\label{Eq:SpatialModeOverlap}
\end{equation}
Because of the opposite parity of the down-converted modes $H$ and $V$, the pump mode needs to be odd for the integral in Eq.~(\ref{Eq:SpatialModeOverlap}) to be non-zero. The most natural candidate is $10_P$. Pump in this mode could be prepared by filtering a laser beam through an auxiliary waveguide, or by using the output of a sum-frequency generation process, which is highly selective in transverse modes for fixed frequencies of fundamental beams \cite{MacHulka2013, Karpinski2009, Roelofs1994}. As shown in Fig.~\ref{Fig:CouplingEfficiencies10}, depicting coupling efficiencies
$|\alpha_{10_{P} \rightarrow {ij}_{H} {i'j'}_{V}}|^2$ for different pairs of down-converted modes, the choice of the $10_P$ mode has an additional advantage of exhibiting the strongest coupling for two processes under consideration. However, two issues remain. A careful inspection of Fig.~\ref{Fig:CouplingEfficiencies10} shows that the overlaps $|\alpha_{10_{P} \rightarrow {00}_{H}  {10}_{V}}|^2$ and $|\alpha_{10_{P} \rightarrow {10}_{H}  {00}_{V}}|^2$ are slightly different, which may prevent generation of a maximally entangled state. Furthermore, although the efficiencies of other processes are suppressed, they remain non-negligible. To address these two issues we will exploit the spectral degree of freedom which also needs to be considered to ensure spectral indistinguishability of the two components in the generated entangled state.

The spectral characteristics of down-converted photons depends on the phase matching function, obtained by integrating along the propagation direction the longitudinal phase factors of the three waves coupled through the non-linear interaction. In a multimode waveguide, the phase matching function depends on the specific triplet of $P$, $H$, and $V$ modes involved in the process and takes the form
\begin{multline}
\phi_{lm_{P} \rightarrow {ij}_{H} {i'j'}_{V}}(\omega_H,\omega_V)
=  \frac{L}{2} \text{sinc} \left[ \frac{L}{2} \left( \vphantom{\frac{L}{2}} k_P^{lm}(\omega_H+\omega_V) \right. \right. \\
\left. \left.
 - k^{ij}_{H} (\omega_H) - k^{i'j'}_{V}(\omega_V) - \frac{2\pi p}{\Lambda} \right) \right].
 \label{Eq:SpectralPhaseMatching}
\end{multline}
Here $\omega_H$ and $\omega_V$ are the frequencies of the down-converted $H$ and $V$ photons, $L$ is the length of the medium and $k_\mu^{ij}(\omega)$ describe the dependence of the wave number on the frequency $\omega$ for the fields $\mu=P, H, V$. The additive term $2\pi p/\Lambda$ is contributed by the quasi-phase matching condition for the order $p$ and the poling period equal to $\Lambda$. 

\begin{figure}[t]
\centering
\includegraphics[width=7.5cm]{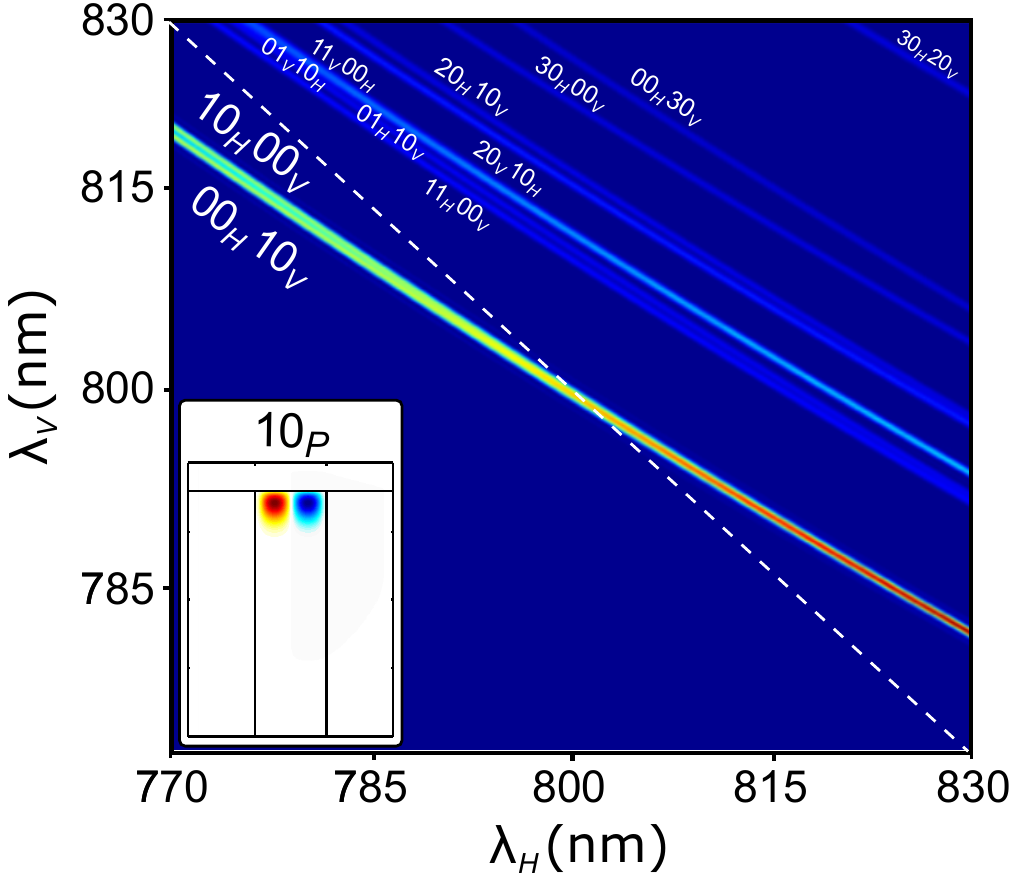}
\caption{Phase matching bands of several first modal processes multiplied by their respective efficiencies for the pump coupled in $10_P$ mode, along with its transversal field profile (inset).  The energy conservation condition curve for the monochromatic pump $\lambda_{P} = 400~\mathrm{nm}$ has been represented by a white dashed line. For a better visualization of individual phase matching bands here we assumed crystal length $L = 4~\mathrm{mm}$.}
\label{Fig:PhaseMatchingBands}
\end{figure}

Intermodal dispersion makes the wave numbers $k_\mu^{ij}(\omega)$ decrease for higher transverse modes. This has significant effect on the phase matching function, as illustrated in Fig.~\ref{Fig:PhaseMatchingBands} for the pump prepared in the $10_P$ mode. It is seen that the phase matching condition $k_P^{10}(\omega_H+\omega_V) = k^{ij}_{H} (\omega_H) + k^{i'j'}_{V}(\omega_V)$ is satisfied along different curves in the plane spanned by wavelengths $\lambda_H = 2\pi c/ \omega_{H}$ and $\lambda_V = 2\pi c/ \omega_{V}$, and that for realistic waveguide length the phase matching bands centered around these curves are usually well separated. Substantial overlap is present for two interactions required to generate an entangled state of two spatial qubits. Such an overlap is needed to ensure spectral indistinguishability of the produced photons that warrants high-visibility of two-photon correlations.
It will be convenient to label the two processes of interest with single digits
\begin{align}
1: & \qquad 10_P \rightarrow 00_H 10_V \nonumber \\
2: & \qquad 10_P \rightarrow 10_H  00_V,
\label{Eq:SimplerLabels}
\end{align}
that will be used in the following as the indices for the respective quantities, such as probability amplitudes.

In the example shown in Fig.~\ref{Fig:PhaseMatchingBands} and further numerical calculations, the poling period of $\Lambda = \mathrm{7.8~\mu m}$ has been selected to maximize the spectral overlap of the two concurrent down-conversion processes specified in Eq.~(\ref{Eq:SimplerLabels}) that are involved in spatial entanglement generation for the waveguide pumped using a 400 nm continuous-wave laser. The corresponding quasi-phase matching order is $p=1$.


\section{Two-photon wave function}
\label{Sec:TwoPhotonWaveFunction}

Let us now consider the complete two-photon wave function including both the spatial and the spectral degrees of freedom.
Owing to energy conservation in the down-conversion process, the sum of the frequencies $\omega_H$ and $\omega_V$ of the generated horizontal and vertical photons must be equal to the frequency of the pump photon $\omega_P = \omega_H + \omega_V$. For cw pumping, this defines a curve in the plane of Fig.~\ref{Fig:PhaseMatchingBands} running at approximately 45$^\circ$  with respect to the graph axes which defines actual frequencies of generated photons (presented as white dashed line). Because the phase matching condition lines are oriented at different angles than 45$^\circ$, the pump frequency constraint introduces separation between the two processes of interest and other combinations of down-conversion spatial modes. Consequently, the latter can be removed using coarse spectral filtering. The feasibility of this approach has been demonstrated for the down-conversion process between fundamental modes, $00_P \rightarrow 00_H 00_V$ in recent experiments with PPKTP waveguides \cite{Karpinski2012,Jachura2014}.

\begin{figure}[b]

\centering
\includegraphics[width=8cm]{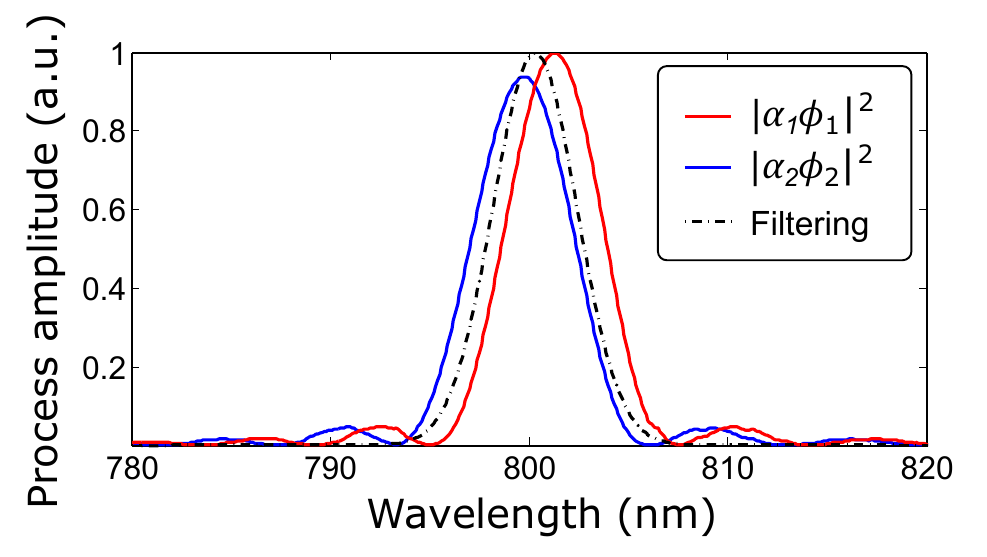}
\caption{Spectra of two concurrent modal processes $10_P \rightarrow 00_H 10_V$ (blue line), $10_P \rightarrow 10_H 00_V$ (red line) for $\lambda_{P} = 400~\mathrm{nm}$ complemented by an exemplary spectral filtering profile. }
\label{Fig:ProcessSpectra}
\end{figure}

Assuming a monochromatic cw pump, the spectral amplitudes of the generated photons can be parameterized with the frequency $\omega$ of the $H$ photon and written as
\begin{align}
\phi_1(\omega)  & = \phi_{10_P \rightarrow 00_H 10_V}(\omega,\omega_P - \omega) \nonumber \\
\phi_2(\omega)  & = \phi_{10_P \rightarrow 10_H 00_V}(\omega,\omega_P - \omega)
\end{align}
In Fig.~\ref{Fig:ProcessSpectra} we depict the effective spectra for the two processes given by expressions $| \alpha_i \phi_i(\omega)|^2$, $i=1,2$, where $\alpha_i$ are spatial overlaps defined in Eq.~(\ref{Eq:SpatialModeOverlap}) labelled using notation introduced in Eq.~(\ref{Eq:SimplerLabels}). It is seen that the produced photon pairs are partially distinguishable, and furthermore the production rates are slightly different. These two detrimental effects can be dealt with by introducing narrowband spectral filtering, shown  in Fig.~\ref{Fig:ProcessSpectra} with a dash-dotted line. Adjusting the central frequency $\omega_0$ of the filter can concurrently reduce distinguishability and equalize contributions from the two processes of interest. We will assume a gaussian profile for the filter with unit transmission at the maximum
\begin{equation}
f(\omega) = \exp[-(\omega-\omega_0)^2/2\sigma^2]
\label{Eq:SpectralFilter}
\end{equation}
where $\sigma$ characterizes the filter bandwidth. This filter function multiplies the spectral amplitudes $\alpha_i \phi_i(\omega)$. It will be convenient to denote overall production rates of filtered photons for processes $i=1,2$ as
\begin{equation}
R_i = \int d\omega \, \left| \alpha_i f(\omega) \phi_i(\omega) \right|^2,
\end{equation}
and to introduce normalized spectral amplitudes
\begin{equation}
\psi_i (\omega)  = \frac{1}{\sqrt{R_i}} \alpha_i f(\omega) \phi_i(\omega).
\end{equation}
The complete two-photon wave function in the position representation, parameterized with two-dimensional transverse positions ${\bf r}_H$ and ${\bf r}_V$ of the respective photons is given for cw pumping by the expression
\begin{multline}
\Psi({\bf r}_H, {\bf r}_V ; \omega) = \sqrt{\frac{R_1}{R_1+R_2}} \psi_1(\omega) u_H^{00}({\bf r}_H) u_V^{10}({\bf r}_V) \\
+ \sqrt{\frac{R_2}{R_1+R_2}} \psi_2(\omega) u_H^{10}({\bf r}_H) u_V^{00}({\bf r}_V)
\label{Eq:Psiromega}
\end{multline}
The overall production rate of photon pairs after filtering is $R_1 + R_2$.

\begin{figure}[tb]
\centering
\includegraphics[width=7.5cm]{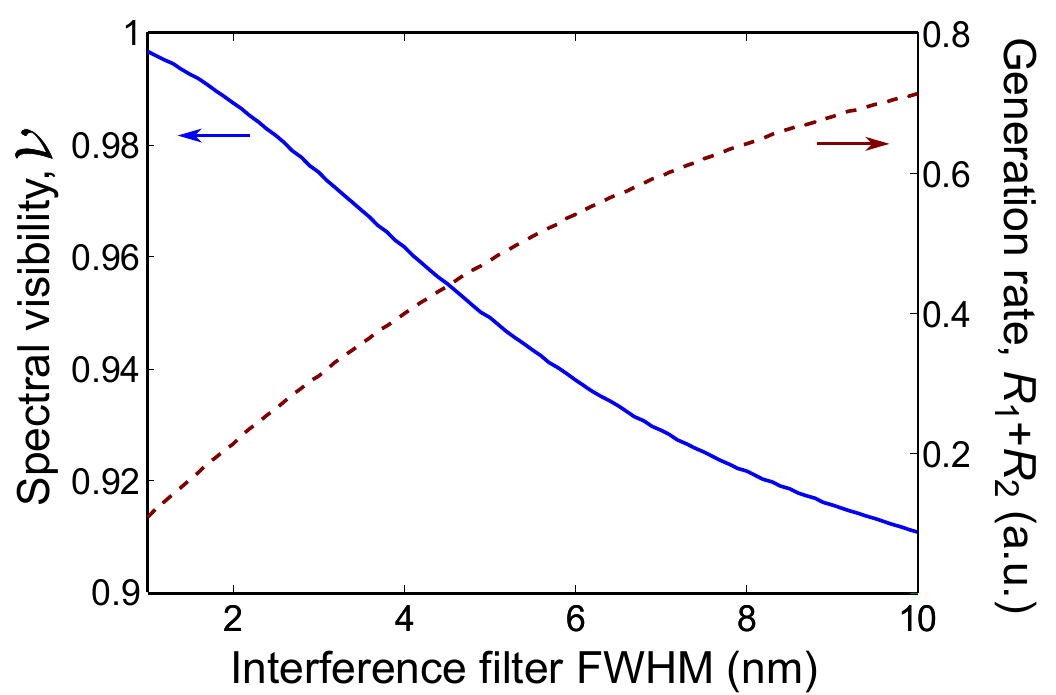}
\caption{Spectral visibility ${\cal V}$ (solid line, left scale) and the overall generation rate $R_1 + R_2$ (dashed line, right scale) as a function of the bandwidth of the interference filter.}
\label{Fig:VisandGenRate}
\end{figure}

When the spectral degree of freedom is traced out, coherence between the two terms in the superposition can be characterized
with the visibility parameter
\begin{equation}
{\cal V} =  \int d\omega \, [\psi_1(\omega)]^\ast \psi_2(\omega).
\end{equation}
Fig.~\ref{Fig:VisandGenRate} presents as a function of the filter bandwidth the absolute value of the visibility parameter $|{\cal V}|$ as well as the pair generation rate which has been normalized to unity in the absence of filtering.
The central frequency of the filter $\omega_0$ in  Eq.~(\ref{Eq:SpectralFilter}) is chosen for all the bandwidths such that $R_1=R_2$. It is seen that that the state can be brought close to the maximally entangled form while the total production rate $R_1+R_2$ remains at a reasonable level.

\section{Qubit characterization}
\label{Sec:Characterization}

The spatial qubits are defined by selecting two specific transverse modes for the continuous position variable. Experimental identification of these modes may be required e.g.\ to ensure mode matching with other integrated optics devices. In this section we will discuss how to perform such an identification using the inverting interferometer shown in Fig.~\ref{Fig:Interferometer}.

For simplicity, we will consider only the $y$ coordinate, with respect to which the qubit basis modes exhibit opposite parity. The single-photon density matrix for the spatial coordinate $y$ in the case of the horizontally polarized photon $H$ reads
\begin{multline}
\varrho_H(y, y') =  \int dz_H \int dy_V \int dz_V \int d\omega \\
 \Psi(  y  , z_H, y_V, z_V ; \omega) \Psi^\ast( y' , z_H,  y_V, z_V  ; \omega) \\
= \frac{1}{R_1+R_2} \int dz \, [R_1 u_{H}^{00}(y,z)  u_{H}^{00}(y',z) \\
+ R_2 u_{H}^{10}(y,z)  u_{H}^{10}(y',z)]
\label{Eq:varrhoHxx'}
\end{multline}
where the second expression uses the explicit form of the two-photon wave function given in Eq.~(\ref{Eq:Psiromega}) and the orthogonality of the mode functions $u_V^{00}({\bf r}_V)$ and $u_V^{10}({\bf r}_V)$. The density matrix $\varrho_V(y,y')$ for the photon $V$ is defined analogously.

 \begin{figure}[tb]
\centering
\includegraphics[width=9cm]{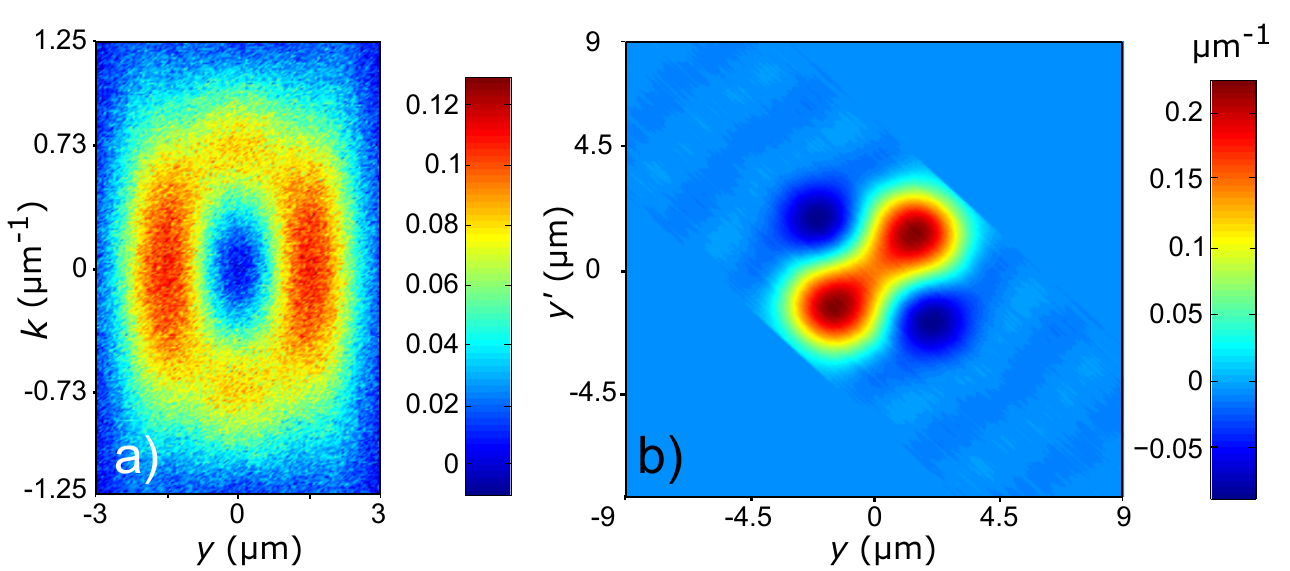}
\caption{(a) Numerical simulation of $H$-polarized photon Wigner function for nearly maximally spatially entangled state measured on a grid of $3 \times 10^{5}$ experimental points with a signal-to-noise ratio of SNR = 12.5. (b) Density matrix of $H$-polarized photon obtained from the Wigner function by applying the transformation defined in Eq.~\ref{Eq: RHOFROMWIGNER}}.
\label{Fig:wignerandrho}
\end{figure}

Displacing and tilting the input beam before entering the inverting interferometer and counting photons at its outputs allows to measure point-by-point the spatial Wigner function \cite{Mukamel2003, Smith2005a}, which is an equivalent representation of the density matrix related through the formula \cite{Schleich2011}
\begin{multline}
W_H(y, k) = \frac{1}{\pi} \int d\xi \,  e^{-2i k \xi} \varrho_H(y+\xi,y-\xi)
\label{Eq:SinglePhotonWigner}
\end{multline}
The arguments of the Wigner function parameterizing the phase space can be scanned directly by steering the input beam before the interferometer entrance: the position $y$ is defined by the displacement with respect to the interferometer axis, whereas the wave number $k$ is proportional to the tilt $\theta$ of the input beam. In Fig.~\ref{Fig:wignerandrho}(a) we present a numerical simulation of such a scan on a $120 \times 250$ grid. The simulation included additive Gaussian noise at each phase space point with standard deviation equal to $0.08$ of the maximum absolute value of the Wigner function. This yields the signal-to-noise ratio of 12.5 defined as the ratio of the maximum Wigner function value to the standard deviation of additive Gaussian noise. In numerical calculations we assumed a two-photon state $\Psi({\bf r}_H, {\bf r}_V ; \omega)$ defined in Eq.~(\ref{Eq:Psiromega}) with component weights $R_{1}/(R_{1}+R_{2}) = 0.4933$ and $R_{2}/(R_{1}+R_{2}) = 0.5067$. These are the most balanced values obtained for numerical optimization over the poling period $\Lambda$ when no spectral filtering is applied. Removing the spectral filter increases the photon rate, hence reducing the duration of the phase space scan. 

\begin{figure}[tb]
\centering
\includegraphics[width=6cm]{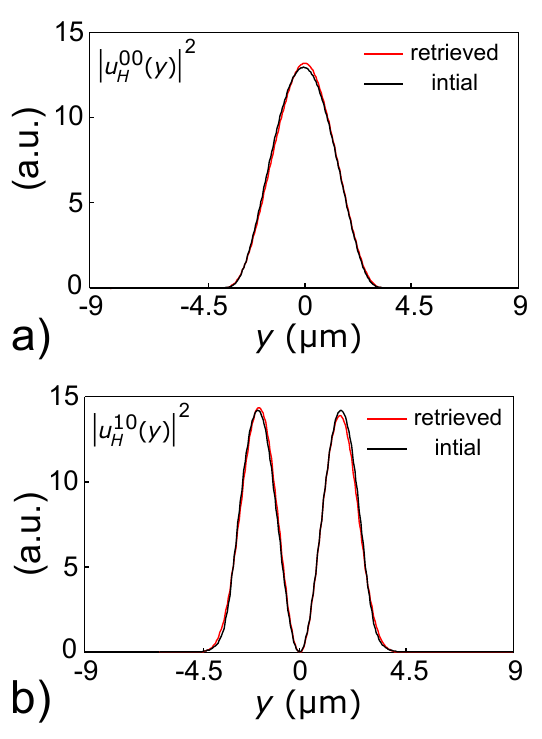}
\caption{Spatial profiles of (a) $u^{00}_{H}(x)$ and (b) $u^{10}_{H}(x)$ basis modes reconstructed from numerically simulated noisy Wigner function compared with their initially assumed counterparts.}
\label{Fig:reconstructedmodes}
\end{figure}

As suggested by the second explicit expression for $\varrho_H(y,y')$ in Eq.~(\ref{Eq:varrhoHxx'}), diagonalization of the single-photon density matrix should yield explicitly the basis modes. To implement this procedure, we used the simulated noisy Wigner function to calculate the density matrix on a discrete grid of $360 \times 360$ points extending over the $-9~\mathrm{\mu m} \le y, y' \le \mathrm{9~\mu m}$ range with the help of an inverse formula to Eq.~(\ref{Eq:SinglePhotonWigner}):
\begin{equation}
\varrho_H (y,y') = \int dk \, e^{ik(y-y')} W_H\bigl( (y+y')/2, k \bigr).
\label{Eq: RHOFROMWIGNER}
\end{equation}
Subsequently, using a standard diagonalization algorithm available in the MATLAB computing environment we decomposed the reconstructed density matrix into eigenvalues $w_n$ and eigenvectors $u_n(y)$:
\begin{equation}
\varrho_H(y, y') = \sum_{n} w_n u_n(y) u_n^\ast(y').
\end{equation}
In Fig.~\ref{Fig:reconstructedmodes} we compare the squared absolute values of eigenvectors $|u_n(y)|^2$ corresponding to two highest eigenvalues with marginal intensity distributions $\int dz |u_H^{ij}(y,z)|^2$, $ij=00,10$ of the modes used in simulations. The agreement is very good despite the statistical noise included in the simulated experimental data. It is noteworthy that these results have been obtained with the second spatial coordinate $z$ traced out in the reconstruction procedure. This shows that for the model waveguide modes considered here the $y$ and $z$ coordinates can be treated as effectively uncorrelated, which greatly simplifies practical characterization of the qubit modes. We found that for noisy input data, slightly unequal rates $R_1$ and $R_2$ stabilize the result of the diagonalization procedure. This is easily understood, as for a maximally entangled state, when $R_1=R_2$, an arbitrary superposition of the basis modes is an eigenvector of the single-photon density matrix.

\section{Testing Bell's inequalities}
\label{Sec:Bell}

One of the striking consequences of entanglement is the violation of Bell's inequalities, which rules out a broad class of theories alternative to quantum mechanics based on local hidden variable theories. We will now describe a method to verify the spatial entanglement of the two-photon state produced the waveguide source by measuring and correlating displaced mode parities using an inverting interferometer shown in Fig.~\ref{Fig:Interferometer}.

In a commonly used scenario for testing Bell's inequalities described by Clauser, Horne, Shimony, and Holt (CHSH), two separated parties perform measurements with dichotomic outcomes $\pm 1$. In each experimental run, they choose randomly and independently between two settings of their measuring apparatuses, $\zeta_H$ or $\zeta_H'$ for one party, and $\zeta_V$ or $\zeta_V'$ for the other party. After a series of measurements, calculating the average product of outcomes on both sides for each pair of settings yields four correlation functions $C(\zeta_H,\zeta_V)$, $C(\zeta_H,\zeta_V')$, $C(\zeta_H',\zeta_V)$, and $C(\zeta_H',\zeta_V')$. These quantities are used to evaluate the combination
\begin{multline}
{\cal B} = C(\zeta_H,\zeta_V) + C(\zeta_H,\zeta_V') + C(\zeta_H',\zeta_V) - C(\zeta_H',\zeta_V')
\end{multline}
whose absolute value for local hidden variable theories cannot exceed two, $-2 \le {\cal B} \le 2$.

\begin{figure}[t]
\centering
\includegraphics[width=7.5cm]{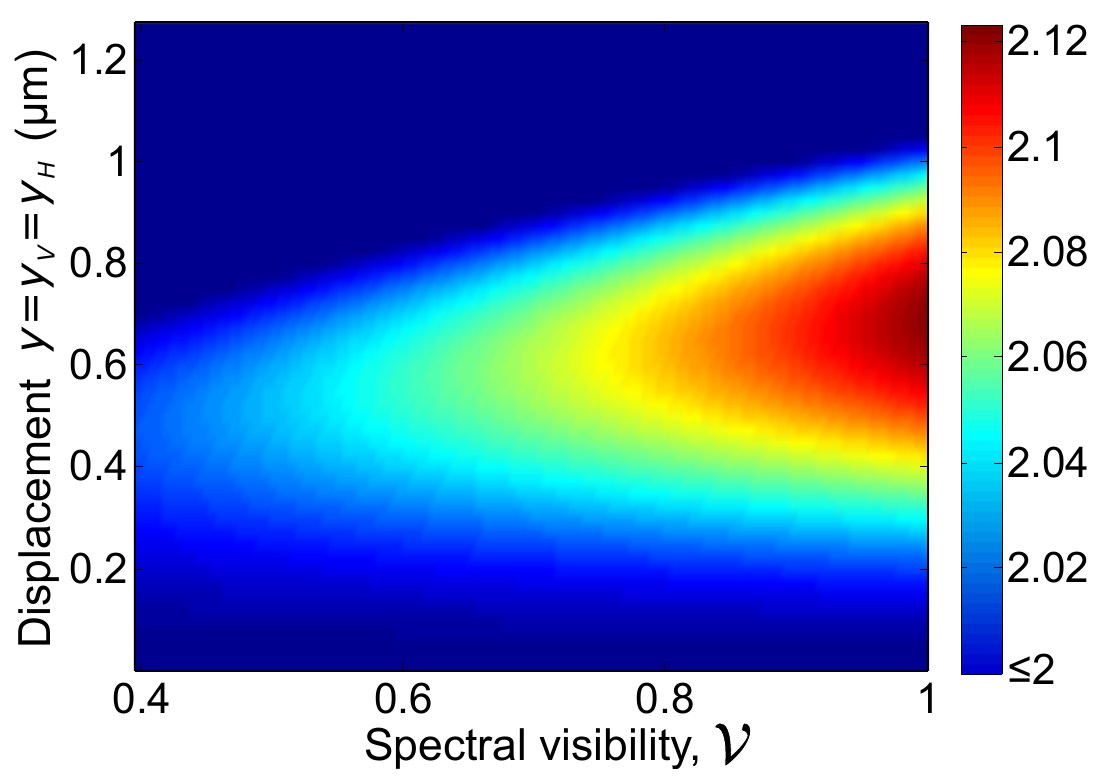}
\caption{Absolute value of CHSH combination with respect to spectral visibility and displacements $\xi=\xi_H =\xi_V$, $\xi'_H=\xi'_V=0$.}
\label{Fig:BellViolation}
\end{figure}

In the case of a pair of spatially entangled photons prepared in a state $\Psi({\bf r}_H, {\bf r}_V ; \omega)$ described by Eq.~(\ref{Eq:Psiromega}), the dichotomic measurement can be implemented using the inverting interferometer with the binary $\pm 1$ result corresponding to the parity of the mode in which a photon has been detected at the output. As the controllable setting of the measuring apparatuses, the parties can use a lateral displacement of the input beam in a direction perpendicular to the symmetry axis of the inverting interferometer. A straightforward calculation shows that the correlation function measured in this case is given by the following expression:
\begin{multline}
C(\zeta_H,\zeta_V) = \int d^2 {\bf r}_H \int d^2 {\bf r}_V \int d\omega \\
\Psi(  \zeta_H + y_H , z_H, \zeta_V + y_V, z_V ; \omega)
 \\
\times \Psi^\ast( \zeta_H - y_H , z_H, \zeta_V - y_V, z_V  ; \omega)
\end{multline}
where integrations in the transverse plane have been parameterized with ${\bf r}_H = (y_H, z_H)$ and ${\bf r}_V = (y_V, z_V)$.
It is worth noting that for zero displacements we have $C(0,0)=-1$, as the modes of the two produced photons always have opposite parities. The maximum value of the CHSH combination optimized over $\zeta=\zeta_H =\zeta_V$, assuming that the second pair of displacements is zero, $\zeta'_H=\zeta'_V=0$, is shown in Fig.~\ref{Fig:BellViolation}.

A stronger violation of the CHSH inequality can be demonstrated using deterministic single-qubit gates, which would facilitate projections onto arbitrary superpositions of the basis states in strict analogy with spin-$1/2$ measurements \cite{Bharadwaj2015}. In this case the combination ${\cal B}$ could approach the maximum value permitted by quantum mechanics equal to $2\sqrt{2}\approx 2.828$, assuming that the two components of the entangled state are balanced and the visibility parameter ${\cal V}$ is sufficiently close to one.

\section{Conclusions}
\label{Sec:Conclusions}

In this paper we analyzed theoretically generation of an entangled state of two spatial qubits, each spanned by a pair of transverse modes in a multimode nonlinear waveguide. We showed that despite the presence of higher transverse modes, it is possible to remove their contribution to the down-conversion process by coarse filtering. This approach is facilitated by the effect of intermodal dispersion, which correlates pairs of individual transverse modes with wavelengths at which down-conversion can occur. The properties of the generated state can be fine-tuned by additional narrowband spectral filtering.

We also considered the inverting interferometer, which detects the transverse spatial parity of the input beam, as a tool to characterize the generated state. We showed that a single-mode Wigner function measured via a phase space scan realized by displacing and tilting the input beam can be used to identify the basis modes spanning the spatial qubit. Interestingly, for realistic parameters of the multimode waveguide, it is sufficient to carry out this procedure in only one spatial coordinate. Simultaneous detection of both photons with independently controlled spatial displacement provides a scheme to verify generated entanglement via a violation of the standard Clauser-Horne-Shimony-Holt inequality.

\section*{ACKNOWLEDGEMENTS}

This work is part of the project "Quantum Optical Communication Systems" carried out within the TEAM programme of the Foundation for Polish Science co-financed by the European Union under the European Regional Development Fund.

\newpage

\bibliographystyle{apsrev4-1}



\end{document}